\documentclass[fleqn,twoside]{article}
\usepackage{espcrc2}
\usepackage{graphicx}
\usepackage[figuresright]{rotating}

\newcommand{\AmS}{{\protect\the\textfont2
  	A\kern-.1667em\lower.5ex\hbox{M}\kern-.125emS}}

\title{Mass spectrum of N* and source optimization}

\author{LHP Collaboration:
	S. Basak\address[umd]{Department of Physics, 
        University of Maryland, College Park, MD 20742, USA},
	R. Edwards\address[jlab]{Thomas Jefferson National Accelerator 
	Facility, Newport News, VA 23606, USA},
	R. Fiebig\address{Physics Department, Florida International University,
	Miami, FL 33199, USA},
	G. Fleming\addressmark[jlab],
	U. M. Heller\address{American Physical Society, One Research Road,
	Ridge, NY 11961-9000, USA},
	C. Morningstar\address{Department of Physics, Carnegie Mellon
	University, Pittsburgh, PA 15213, USA},
	D. Richards\addressmark[jlab],
	I. Sato\addressmark[umd]\thanks{Presented by I. Sato}, and
	S. Wallace\addressmark[umd].}
       
\begin{document}

\begin{abstract}

We have computed correlation functions for nucleons and extracted the masses for 
positive- and negative-parities.  Use of group theory
plays an important role in obtaining sources that have good overlap
to higher spin states and minimum contamination from unwanted states.
In the simulation three distinct sources and corresponding sinks that transform
according to the $G_1$ irreducible representations are tested 
and used to form matrices of correlation functions.
Diagonalizations give us clear mass
splittings between low-lying states and excited states for both parities.

\end{abstract}

\maketitle

\section{INTRODUCTION}

Lattice QCD calculations of baryon masses are based on 
the analysis of 2-point correlation functions, defined by
\begin{equation}
C_{ij}(t,t_0)=\sum_{\bf x} \left< 0 \right| B_i ({\bf x},t) \overline{B}_j (0,t_0)
\left| 0 \right>,
\label{eq:101}
\end{equation}
where $B_i$ are color-singlet baryon source fields.
Summation over ${\bf x}$ projects to zero total momentum.  
For large $t$ the contribution from the lowest mass eigenstate dominates.
An effective mass is calculated from 
 $M_{\rm eff}=ln \left( {C(t,t_0) \over C(t+1,t_0)} \right)$.

Ideal sources maximally overlap the states of interest:
$\left< n \right| \overline{B}_i (t_0) \left| 0 \right> \sim \delta_{ni}$,
where $n$ is an intermediate state.
By using the orthogonality relation of basis vectors spanning irreducible 
representations (IRs) of the cubic group, one can attempt to 
realize this relation by taking $B_i$ to be a linear combination of 
source operators that have a good overlap with state $n$. 

There are 48 discrete spatial rotations that conserve the cubic symmetry of 
spin $1/2$ objects, and they form a group, called 
%There are 24 discrete spatial rotations that leave a lattice invariant, 
%and they form a group, called, ``octahedral group''.  
%With spin degree of freedom, the number of rotations that leave a lattice
%invariant becomes 48.  The group consisting of 
%these 48 elements is named 
``double covered octahedral group'', 
or  $ ^2 \cal O$.  
The reduction of the group $ ^2 \cal O$ falls into three spinorial IRs: 
$ G_1,G_2$, and $H$ with respective 
dimensions 2, 2, and 4.  
Table~\ref{table:1} 
lists the inverse relation of the reduction of 
SU(2) to $ ^2 \cal O$~\cite{Johnson82}.
\begin{table}[h]
\caption{Correlation of IRs and total angular momentum states.}
\label{table:1}
\newcommand{\m}{\hphantom{$-$}}
\newcommand{\cc}[1]{\multicolumn{1}{c}{#1}}
\renewcommand{\tabcolsep}{2pc} % enlarge column spacing
\renewcommand{\arraystretch}{1.2} % enlarge line spacing
\begin{tabular}{@{}ll}
\hline
IRs           & $j$                      \\
\hline
$G_1$   & $1/2, 7/2, 9/2, 11/2, ...$ \\
$G_2$   & $5/2, 7/2, 11/2, ...$       \\
$H$     & $3/2, 5/2, 7/2, 9/2, ...$    \\
\hline
\end{tabular}\\[2pt]
\end{table}
Fermionic operators on a cubic lattice 
can be written in terms of basis operators that span IRs of $ ^2 \cal O$. 
Since lattice calculations preserve cubic symmetry, 
different IRs do not admix.
However, each IR is an admixture of different total 
angular momenta.
Total angular momentum is not a ``good''
 quantum number on the lattice because
continuous rotational symmetry is broken.
It is expected that higher spin states have higher masses, therefore
within a given IR the lowest spin state dominates at large $t$.

\section{NONLOCAL SOURCES}

A general expression of three-quark baryon operator would be written as
\begin{eqnarray}
B_\mu ({\bf x})=\sum A({\bf x}, {\bf l}_1, {\bf l}_2, 
{\bf l}_3, \mu, \alpha, \beta, \gamma) 
\epsilon^{abc} 
\nonumber \\
U^{a, a'}_1 q^{a'}_\alpha ({\bf l}_1) 
U^{b, b'}_2 q^{b'}_\beta ({\bf l}_2)
U^{c, c'}_3 q^{c'}_\gamma ({\bf l}_3),
\label{eq:21}
\end{eqnarray}
where $\alpha, \beta$, and $\gamma$ are Dirac indices, 
$a$, $b$, $c$, $a'$, $b'$, and $c'$ are 
color indices, and ${\bf l}_1, {\bf l}_2$, and ${\bf l}_3$ are spatial 
displacement vectors.  
$U^{a,a'}_1$ is the gauge link connecting the spatial path 
from ${\bf x}$ to ${\bf l}_1$.  The coefficients tie together a set
of spatial displacement vectors ${\bf l}_1, {\bf l}_2$, and ${\bf l}_3$, 
and Dirac indices $\alpha, \beta$, and $\gamma$ in such a way 
as to yield gauge invariant operators that transform according to IRs of 
$ ^2{\cal O}$ and definite isospin.

In order to have sources that are commensurate with the size of baryons,
it is appropriate to smear the sources, using for example, Gaussian
smearing,
\begin{equation}
F_0 = \left( 1- {\sigma^2 D^2 \over 4N} \right)^N,
\label{eq:103}
\end{equation}
where $\sigma$ is the smearing width, $N$ is an integer, 
and $D^2$ is the gauge-covariant
three-dimensional Laplacian operator.  For different radial distributions, 
additional powers of Laplacian operators are used, e.g., 
$F_2 = D^2 F_0$, $F_4 = D^2 F_2$, and so on.

The simplest nonlocal operator has ${\bf l}_1 = {\bf l}_2 = {\bf x}$, 
${\bf l}_3 = {\bf x} + a{\bf \hat{e}}_1$.
Applying spatial rotations of the cubic group 
to this operator leads to $6\times 8=48$ dimensional representations,
where six is for the directions of $a {\bf \hat{e}}_1$ and eight for the three
spin degrees of freedom.  Reduction of these
representations results in 6 $G_1$'s, 2 $G_2$'s, and 8 $H$'s.
From these ``one-link'' constructions, we used both positive- and negative-parity 
nucleon sources from $G_1$ basis in the simulation.

More complicated sources have been constructed by choosing the initial 
operator as, for instance, 
${\bf l}_1 = {\bf l}_2 = {\bf x}$, ${\bf l}_3 = {\bf x} + a{\bf \hat{e}}_1 
+ a{\bf \hat{e}}_2$ in Eq.~(\ref{eq:21}).

\section{SIMULATION DETAILS AND PRELIMINARY RESULTS}

For a test of these sources, we formed two 
$3\times 3$ matrices of correlation functions with
positive-parity and negative-parity, by using projection operators
$1 \pm \gamma_4$.
All operators transform according to $G_1$ IRs.
Isospin is selected to be $I=I_z=1/2$.  The operators used in the 
simulation are shown below,
\begin{eqnarray}
s^{(\pm)}_{1/2} \equiv \left[ d (C\gamma_5) u \right]
(1 \pm \gamma_4 ) (1+\gamma_3 \gamma_5) u,
\label{eq:1}
\\
D^2 s^{(\pm)}_{1/2} \equiv \left[ d (C\gamma_5) u \right]
(1 \pm \gamma_4) (1+\gamma_3 \gamma_5) D^2 u,
\label{eq:2}
\\
p^{(\pm)}_{1/2} \equiv \left[ d (C\gamma_5) u \right]
(1 \mp \gamma_4) (1+\gamma_3 \gamma_5) D_+ u
\nonumber
\\
+ \left[ d (C\gamma_5) u \right]
(1 \mp \gamma_4) (1-\gamma_3 \gamma_5) D_z u,
\label{eq:3}
\end{eqnarray}
where $D_i$ is a covariant derivative and $D_+ = D_x + i D_y$.  Color and 
Dirac indices are implicit here.
The upper sign is for positive-parity and the lower sign is for 
negative-parity.  All quark fields are smeared using Eq.~(\ref{eq:103}).
Equation~(\ref{eq:1}) is the widely used $G_1$ irreducible source,
which mainly couples to the ground state of the nucleon: we call it 
``$s^{(\pm)}_{1/2}$''
even though from Table~\ref{table:1} one sees that there can be admixtures 
with spin 7/2 or higher.
Equation~(\ref{eq:2}) is s-wave with
an additional Laplacian operator acting on the third quark.  
Equation~(\ref{eq:3}) is constructed 
from one-link sources as described above.  
The derivative operators and spin-projection operators in Eq.~(\ref{eq:3})
that act on the third quark can be rewritten as $D_+ u_{-1/2} + D_z u_{1/2}$.
This corresponds to the Clebsch-Gordan formula for a $p_{1/2}$ state,
i.e., 
\begin{equation}
p_{1/2}=
-\sqrt{1 \over 3}Y_{11} \chi_{-1/2}
+\sqrt{2 \over 3}Y_{10} \chi_{1/2},
\end{equation}
where $Y_{lm}$ represents a discretized version of a spherical harmonic, 
and $\chi_{\pm 1/2}$ represent up and down two-component spinors.  
Because of this correspondence,
Eq.~(\ref{eq:3}) is called the ``$p^{(\pm)}_{1/2}$'' source.

We use $16^3 \times 32$ isotropic lattice with periodic boundary conditions.
The smearing width that is used in Eq.~(\ref{eq:103}) is $\sigma = 4.5$, 
with $N=50$.  
We use 221 configurations\footnote{Configurations are from the NERSC lattice
archive.} generated from the isotropic Wilson gauge action at $\beta=6.0$ 
and the Wilson fermion action with $\kappa=0.1550$, which corresponds to pion mass 
about 542 MeV.

\begin{figure}[th]
\includegraphics[angle=270,width=18.5pc]{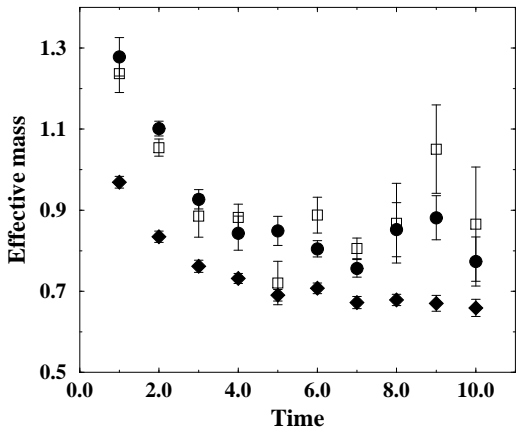}
\caption{Effective masses of positive-parity $G_1$ states,
related to the diagonal correlators mentioned in the text.}
\label{fig:1}
\end{figure}

\begin{figure}[thb]
\includegraphics[angle=270,width=18.5pc]{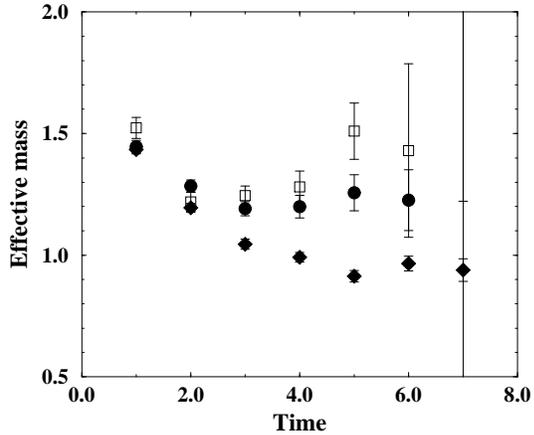}
\caption{Effective masses of negative-parity $G_1$ states.}
\label{fig:2}
\end{figure}

We adopt a variational method to extract effective 
masses~\cite{ukqcd93,michael85,lw90}.
Correlation matrices are diagonalized at a time slice $t_v$.   
This produces three orthogonal eigenvectors, each of the form 
${\bf v} = (v_1, v_2, v_3)$, where $v_i$'s are the coefficients of 
$s^{(\pm)}_{1/2}$, $D^2 s^{(\pm)}_{1/2}$, and $p^{(\pm)}_{1/2}$, respectively.
The resulting eigenvectors are then used to calculate  
``diagonal'' correlators of the form ${\bf v}^T C(t) {\bf v}$ 
at other times.  
For positive-parity the eigenvectors obtained at $t_v=10$ 
give the most clear mass splitting, which is shown in
Fig.~(\ref{fig:1}).  Filled diamonds have ${\bf v} \approx (0.44, 0.57, 0.68)$, 
i.e., each source is almost equally weighted.  
This state reaches a plateau for $5\leq t\leq 10$, 
suggesting that the ground state of the nucleon has been obtained.
Filled circles and open boxes are for other orthogonal
eigenvectors.  The effective masses based on the other orthogonal eigenvectors 
are larger than the ground state mass, indicating that excited
states of positive-parity are being resolved.   

For negative-parity, the eigenvectors at $t_v=6$ give the most clear mass 
splitting, which is shown in Fig.~(\ref{fig:2}).  Filled diamonds have 
${\bf v} = (0.37, 0.56, 0.71)$, where the contribution from the 
$p^{(-)}_{1/2}$ state 
is significant.
The effective mass of this state is about 40\% larger than the 
positive-parity ground state; the corresponding $N^{*}({1/2}^-)$ physical state 
is about 63\% heavier than the nucleon.  
Filled circles and open boxes have larger error bars,
but they indicate larger masses than diamonds.

The present results are preliminary.  Work is under way to improve the chiral 
properties of the action and to use a large set of sources with anisotropic
lattices.

This work was supported by the U.S. National Science Foundation under
Awards PHY-0099450 and PHY-0300065, and by the U.S. Department of Energy 
under contract DE-AC05-84ER40150 and DE-FG02-93ER-40762.

\end{document}